# Sizable spin-transfer torque in Bi/Ni$_{80}$Fe$_{20}$ bilayer film

Masasyuki Matsushima [1], Shinji Miwa [2], Shoya Sakamoto [2], Teruya Shinjo [1], Ryo Ohshima [1], Yuichiro Ando [1], Yuki Fuseya [3] and Masashi Shiraishi [1,*]

1. Department of Electronic Science and Engineering, Kyoto University, Kyoto 615-8510, Japan.
2. The Institute for Solid State Physics, The University of Tokyo, Chiba 277-8581, Japan.
3. Department of Engineering Science, University of Electro-Communications, Chofu, Tokyo 182-8585, Japan.

* Corresponding author: Masashi Shiraishi (shiraishi.masashi.4w@kyoto-u.ac.jp)

**Abstract**

The search for efficient spin conversion in Bi has attracted great attention in spin-orbitronics. In the present work, we employ spin-torque ferromagnetic resonance to investigate spin conversion in Bi/Ni$_{80}$Fe$_{20}$(Py) bilayer films with continuously varying Bi thickness. In contrast with previous studies, sizable spin-transfer torque (i.e., a sizable spin-conversion effect) is observed in Bi/Py bilayer film. Considering the absence of spin conversion in Bi/yttrium-iron-garnet bilayers and the enhancement of spin conversion in Bi-doped Cu, the present results indicate the importance of material combinations to generate substantial spin-conversion effects in Bi.

The spin-orbit interaction (SOI) plays a pivotal role in spin-orbitronics [1], which is a family of modern spintronics enabling study on spin conversion physics [2] and practical applications for spin-torque and spin-Hall devices [3], wherein a mixed state of spin and orbital angular momenta provides rich fundamental physics and broad possibilities for practical spintronics devices. In spin-orbitronics, the current understanding is that the magnitude of the SOI determines, in principle, the efficiency of spin conversion. The magnitude of the SOI has been naïvely understood to be proportional to the fourth power of the atomic number $Z$ [4,5], which means that heavy materials such as Pt, W, Ta and Bi have been considered as candidate materials for efficient spin conversion.

The history of spin-orbitronics is the chronicle of a quest for materials with high spin-conversion efficiency. Platinum (Pt) is the central material in the field, and a number of significant spin conversion effects have been discovered in Pt, for instance, considerably large spin Hall angle (an index of spin-charge interconversion efficiency and defined as the ratio of spin current to charge current densities) of +0.1 by using spin Hall magnetoresistance [6] and +0.08~+0.09 by using spin-torque ferromagnetic resonance (ST-FMR) [7,8], magnetic damping [9], the inverse spin Hall effect (ISHE) [10], a tunable spin Hall effect from the dirty to the super-clean regimes [11], and ionic-gating-induced modulation of the spin Hall conductivity in ultrathin film [12]. Note that the magnitude $\zeta$ of the SOI for the 5d electrons in Pt was calculated to be 46.1 mRy [13,14]. If the SOI of a material is the sole factor that determines the spin-conversion efficiency, Bi ( $\zeta$ ~106.8 mRy for the 6p electrons), which has been exhibiting fruitful condensed matter physics such as such electron fractionalization, bulk superconductivity, large ordinal magnetoresistance, and intriguing empty Dirac valley states [15-18], would exhibit the highest efficiency of all non-radioactive elements [19], and this is the central reason for growing interest in Bi for spin-orbitronics. In fact, significant effort has already been expanded to realize the high spin-conversion efficiency using Bi.

Despite the efforts to realize efficient spin conversion in Bi, long-term controversy remains because the reported spin Hall angle of Bi depends strongly on the measuring techniques used and the sample structures [20-25] and the magnitude is much smaller than those of Pt [5], β-Ta [26] and β-W [27], which contradicts the theoretical prediction that the SOI of Bi should be quite large. In fact, the recent study by Yue *et al.* using the spin Seebeck effect claimed a negligible ISHE in rhombohedral (111) Bi grown on yttrium-iron-garnet (YIG) [23]. Meanwhile, studies on spin conversion in Bi film using dynamical spin pumping reported a wide variety of sizable spin Hall angles (−0.071~+0.02) [20,21]. In particular, a surprising result is the sign inversion (from negative to positive) of the spin Hall angle of Bi that takes place for a Bi thickness of 2 nm [20], the physics of which remains inconclusive. Another notable aspect is that Bi impurities in a light element such as Cu exhibits quite large spin Hall angles (−0.24 [24] and +0.12 [25]), where each study reported a different sign of the spin Hall angle. These two studies exploited spin absorption and mesoscopic Hall-bar methods, respectively, but both used electrical methods.

An additional open question in Bi-based spin-orbitronics is the appearance or absence of the inverse Rashba-Edelstein effect (IREE) in Bi/Ag bilayer film. Although studies have reported appearance of the IREE in Bi(111)/Ag [28,29] by using spin pumping, in NiFe/Bi/Ag by using ST-FMR [30] and in Bi(111)/Ge(111) by combining an optical method and the spin-pumping method [31], a study using the spin Seebeck effect [23] reported absence of the IREE. Consequently, spin conversion in Bi remains puzzling and under debate, so further investigation of spin conversion in Bi-based materials systems is highly desirable to provide a deeper understanding of the underlying physics of spin conversion in Bi because material states (bilayer, trilayer, and impurities in the host material) and measuring methods (dynamical spin pumping, optical spin injection, and the spin Seebeck) can play significant roles in determining the spin conversion in Bi-based materials systems.

Here in this study, we employ ST-FMR [6, 32] for a study of spin conversion in Bi and focus on Bi/Ni$_{80}$Fe$_{20}$(Py) bilayer film with a systematically varied Bi thickness to investigate the possible spin conversion in the film and the possible sign inversion of the spin Hall angle at Bi thickness of 2 nm. ST-FMR is one of the most conventional and established methods to understand spin-conversion physics. Thus, further evidence for understanding spin-conversion physics in Bi can be obtained using ST-FMR. In addition, as aforementioned, the interesting sign inversion of the spin Hall angle of Bi was reported only in the combination of Bi and Py under dynamical spin pumping. Thus, the use of ST-FMR to investigate Bi/Py bilayer enables providing important evidence regarding spin conversion in Bi.

The samples used in this study consisted of Bi($t$ mn)/Py(5 nm) grown by molecular beam epitaxy on a thermally oxidized Si substrate. The thickness $t$ of the Bi was varied from 1.2 to 10 nm by introducing a wedged structure on the same substrate. MgO (5 nm) and SiO$_2$ (5 nm) were deposited on the top of the sample to protect the Bi from oxidation. Figures 1(a) and 1(b) show a schematic illustration of the sample and the measurement circuit, respectively. To measure ST-FMR signals, a Bi/Py sample (20 μm × 130 μm) was fabricated using electron-beam lithography and Ar-ion milling in the signal line of a shorted coplanar waveguide. An rf current with an input power of 20 mW (13 dBm) was applied in the longitudinal direction of the sample with an in-plane external magnetic field $\mu_0 H_{\text{ext}}$ at an angle of $\theta=45°$ and $225°$ with respect to the longitudinal direction (see Fig. 1(b)). All measurements are carried out at room temperature.

The FMR is caused by the Oersted field due to a charge current in the Bi layer. The field-induced FMR spectrum has an antisymmetric component in an output voltage at the resonance field. Meanwhile, a spin current due to the spin Hall effect in the Bi layer is injected into the adjacent Py layer and excites a spin-current-induced FMR, which has a symmetric component in the output voltage. The mixed FMR of these two components can be detected as a dc voltage via the rectification effect [6, 32], which can be expressed as,

$$V = -C[SF_S(H_{\text{ext}}) + AF_A(H_{\text{ext}})], \qquad (1)$$

where $C$ is a coefficient concerning the anisotropic magnetoresistance, $F_S(H_{\text{ext}}) = \Gamma^2/[(H_{\text{ext}} - H_{\text{res}})^2 + \Gamma^2]$ is a symmetric Lorentzian with a resonance field $\mu_0 H_{\text{res}}$ with $\mu_0 \Gamma/2$ being the half width at half maximum of the ST-FMR spectrum, $F_A(H_{\text{ext}}) = F_S(H_{\text{ext}})(H_{\text{ext}} - H_{\text{res}})/\Gamma$ is an antisymmetric Lorentzian. The symmetric component $S$ is relevant to the spin current density $J_s$ generated by the spin Hall effect of Bi, whereas the asymmetric component $A$ is due to the sum of the Oersted field around the Bi wire and the field-like torque (i.e., the spin-transfer torque ascribed to the ferromagnet/nonmagnet interface) and is relevant to the charge current density $J_c$. Thus, the spin-conversion efficiency $\eta$ is relevant to the spin Hall angle and describes the ratio of spin current to charge current densities under the assumption of negligible spin-memory loss [33],

$$\eta = \frac{J_s}{J_c} = \frac{S}{A} \frac{e\mu_0 M_s t_{\text{Py}} t_{\text{Bi}}}{\hbar}\left(1 + \frac{M_{\text{eff}}}{H}\right)^{1/2}, \qquad (2)$$

where $e > 0$ is elementrary charge, $t_{\text{Py}}$ (=5 nm) is the Py thickness, $\hbar$ is the Dirac constant, $\mu_0 M_s$

(=0.72 T) is the saturation magnetization of 5-nm-thick Py [34], $\mu_0 M_{\text{eff}}$ is the effective saturation magnetization, and $\mu_0 H$ is equivalent to the resonance field $\mu_0 H_{\text{res}}$. Based on the aforementioned discussion and assumption, the estimated $\eta$ can be the lower limit of the spin Hall angle.

Figure 2(a) shows ST-FMR spectra from the Bi/Py sample as a function of microwave frequency of 7-14 GHz, when the Bi is 5 nm in thick and the external magnetic field was applied at 45°. The black solid lines in the figure are fits to Eq. (1) for all spectra, which determine the coefficients $S$ and $A$. The spectra are well fitted by the sum of the symmetric and the antisymmetric components. Figure 2(b) shows the frequency as a function of the resonance field. A theoretical fitting using the Kittel equation reproduces the experimental results, which allows $\mu_0 M_{\text{eff}}$ to be estimated as 0.77 T for a Py g-factor of 2.112 [35]. The $\mu_0 M_{\text{eff}}$ of 5 nm-thick Py films was reported to be 0.79 T [34]. Therefore, the $\mu_0 M_{\text{eff}}$ in this study is comparable to those in previous studies. To verify that this experimental scheme functioned as designed, we measured how the full width at half maximum of the ST-FMR spectra, $\mu_0 \Delta H = 2\mu_0 \Gamma$, depends on Bi thickness, with the frequency set to 10 GHz and the external magnetic field set to 45° and 225° (see Fig. 2(c)). $\mu_0 \Delta H$ increased monotonically as a function of Bi thickness at the magnetic fields of both 45° and 225°, which is attributed to the fact that the spin-pumping effect [8] from Py to Bi increases monotonically as a function of Bi thickness, and, as designed, the generated spin-transfer torque modulates the magnetization damping in the Py layer. Indeed, Bi thickness dependence of the Gilbert damping constant exhibits similar dependence, which also supports the spin-pumping effect from Py to Bi (see Supplemental Materials).

Figure 3(a) plots the symmetric and the antisymmetric components ($V_S = CS$ and $V_A = CA$, respectively) as functions of Bi thickness. To avoid an artifact originating from asymmetry in the instrument, we averaged the amplitude of the symmetric and antisymmetric components for 45° and 225°. The spin-conversion efficiency $\eta$ as a function of Bi thickness is estimated to be +0.03~+0.06 (see Fig. 3(b)) over the entire thickness range by measuring $\mu_0 M_{\text{eff}}$ of Bi/Py films with various Bi thicknesses and using the linear extrapolation approach (see Supplemental Materials for details). As mentioned above, the spin-conversion efficiency $\eta$ is the lower limit of the spin Hall angle, and the $\eta$ in this study exceeds the spin Hall angle (ca. +0.02) obtained using spin pumping [20,21]. In a study of the ISHE and IREE in Py/Ag/Bi samples, Jungfleisch *et al.* reported ST-FMR from Py (9 nm)/Bi (4 nm) and observed a sizable IREE upon inserting Ag between the Py and the Bi layers. Nevertheless, no ST-FMR signal was detected in the Py/Bi film (i.e., no Ag layer) [30], which is the only report of the ST-FMR effect in Py/Bi. In the present study, however, sizable ST-FMR signals are detected. Thus, the present study provides another perspective for discussing spin-conversion physics in Bi-based materials systems.

Note also the absence of the sign inversion of the spin Hall angle in Bi at around 2 nm in thickness. As discussed above, although $\eta$ is not the same as the spin Hall angle, the two are closely related. The experimental results in the present study unequivocally indicate the absence of sign inversion in $\eta$, which is equivalent to the absence of the sign inversion in the spin Hall angle in Bi, despite the amplitude of $\eta$ in this study not being precisely identical for the Bi thickness.

The results obtained in this study also reveal the importance of materials combinations in Bi-based spin-orbitronics. When we neglect the difference in spin-injection methods, we find that spin conversion is detected in Bi/Py layered systems (ref. [20-22] and this study), but the efficiency is indeed less than that in Pt, W and Ta. In other studies, spin conversion vanished in Bi/YIG combinations [23]. Note, however, that spin conversion was detected in Bi/YIG upon spin pumping [22], but the conversion efficiency of 0.0001 is quite small. A significant spin-conversion efficiency appears only when Bi is doped into a light material (ca. 1% doping in ref. [24] and $\delta$-doped in ref. [25]), which can correspond to the appearance of weak anti-localization in Bi-doped Si [36], where the doping concentration of Bi was $5\times10^{19}$ cm$^{-3}$ (i.e., merely 0.1% dopant concentration in the Si). Although the physics behind of the appearance of the large spin-conversion efficiency only with the Bi as dopants remains inconclusive, further studies of spin conversion in Bi using different methods to unveil more detailed contributions such as spin memory loss and so on [37] and different material combinations and material states can pave the way for a precise understanding of the spin-conversion physics in Bi.

In summary, spin conversion in Bi/Py bilayer films was studied using ST-FMR. In contrast to the previous studies, we detect sizable spin conversion via ST-FMR from Bi/Py samples with various Bi thickness. In addition, no sign inversion of the spin Hall angle is detected. Considering that spin-conversion effects are very weak in Bi/YIG, this study emphasizes the importance of the material combinations to produce spin-conversion effects in Bi.

Supplemental Materials describe Bi thickness dependences of the Gilbert damping constant and of effective saturation magnetization of the Py.

The data that support the findings of this study are available from the corresponding author upon reasonable request.

This study was supported in part by a Grant-in-Aid for Scientific Research (B), "Control of dynamics spin response in strong spin-orbit coupling systems" (No. 19H01850), a Grant-in-Aid for Scientific Research (A) (No. 18H03880), a Grant-in-Aid for Scientific Research (S),

"Semiconductor spincurrentronics" (No. 16H06330) and the Spintronics Research Network of Japan (Spin-RNJ). M.M. acknowledges support by the program of Research Fellowship for Young Scientists from the Japan Society for the Promotion of Science (JSPS). The data that support the findings of this study are available from the corresponding author upon reasonable request.

**References**

1. A. Manchon, H.C. Koo, J. Nitta, S. M. Frolov and R.A. Duine, Nature Mater. **14**, 871 (2015).

2. Y. Otani, M. Shiraishi, A. Oiwa, E. Saitoh and S. Murakami, Nature Phys. **13**, 829 (2017).

3. T. Chen, R.K. Dumas, A. Eklund, P.K. Muduli, A. Houshang, A.A. Awad, P. Dürrenfeld, G. Malm, A. Rusu and J. Åkerman Proc. IEEE **104**, 1919 (2016).

4. J.D. Bjorken and S.D. Drell, "*Relativistic Quantum Mechanics*", *McGraw-Hill Science Engineering*, 1998.

5. H.L. Wang, C.H. Du, Y. Pu, R. Adur, P.C. Hammel and F.Y. Yang, Phys. Rev. Lett. **112**, 197201 (2014).

6. M. Althammer, S. Meyer, H. Nakayama, M. Schreier, S. Altmannshofer, M. Weiler, H. Huebl, S. Geprägs, M. Opel, R. Gross *et al*., Phys. Rev. B **87**, 224401 (2013).

7. A. Ganguly, K. Kondou, H. Sukegawa, S. Mitani, S. Kasai, Y. Niimi, Y. Otani and A. Barman, Appl. Phys. Lett. **104**, 072405 (2014).

8. L. Liu, T. Moriyama, D.C. Ralph and R.A. Buhrman, Phys. Rev. Lett. **106**, 036601 (2011).

9. K. Ando, S. Takahashi, K. Harii, K. Sasage, J. Ieda, S. Maekawa and E. Saitoh, Phys. Rev. Lett. **101**, 036601 (2008).

10. E. Saitoh, M. Ueda, H. Miyajima and G. Tatara, Appl. Phys. Lett. **88**, 182509 (2006).

11. E. Sagasta, Y. Omori, M. Isasa, M. Gradhand, L.E. Hueso, Y. Niimi, Y. Otani and F. Casanova, Phys. Rev. B **94**, 060412(R) (2016).

12. S. Dushenko, M. Hokazono, K. Nakamura, Y. Ando, T. Shinjo and M. Shraishi, Nat. Commun. **9**, 3118 (2018).

13. D.D. Koelling and B.N. Harmon, J. Phys. C **10**, 3107 (1977).

14. T. Kawai, H. Muranaka, T. Endo, N.D. Dung, Y. Doi, S. Ikeda, T.D. Matsuda, Y. haga, H. Harima, R. Settai and Y. Onuki, J. Phys. Soc. Jpn. **77**, 064717 (2008).

15. K. Behnia, L. Balicas and Y. Kopelevich, Science **317**, 1729 (2007).

16. O. Prakash, A. Kumar, A. Thamizhavel and S. Ramakrishunan, Science **355**, 52 (2017).

17. F.Y. Yang, K. Liu, K. Hong, D.H. Reich, P.C. Searson and C.L. Chien, Science **284**, 1335 (1999).

18. Z. Zhu, J. Wang, H. Zuo, B. Fauque, R.D. McDonald, Y. Fuseya and K. Behnia, Nature Commun. **8**, 15297 (2016).

19. When we extend object materials to radiative one, spin conversion in Uranium was reported in the manuscript, S. Singh, M. Anguera, E. del Barco, R. Springell and C.W. Miller, APL **107**, 232403 (2015).

20. D. Hou, Z. Qiu, K. Harii, Y. Kajiwara, K. Uchida, Y. Fujikawa, H. Nakayama, T. Yoshino, T. An, K. Ando, X. Jin and E. Saitoh, Appl. Phys. Lett. **101**, 042403 (2012).

21. H. Emoto, Y. Ando, E. Shikoh, Y. Fuseya, T. Shinjo and M. Shiraishi, J. Appl. Phys. **115**, 17C507 (2014).

22. H. Emoto, Y. Ando, G. Eguchi, R. Ohshima, E. Shikoh, Y. Fuseya, T. Shinjo and M. Shiraishi, Phys. Rev. B **93**, 174428 (2016).

23. D. Yue, W. Lin, J. Li, X. Jin and C.L. Chien, Phys. Rev. Lett. **121**, 037201 (2018).

24. Y. Niimi, Y. Kawanishi, D. H. Wei, C. Deranlot, H. X. Yang, M. Chshiev, T. Valet, A. Fert, and Y. Otani, Phys. Rev. Lett. **109**, 156602 (2012).

25. C. Chen, D. Tian, H. Zhou, D. Hou and X. Jin, Phys. Rev. Lett. **122**, 016804 (2019).

26. L. Liu, C-F. Pai, Y. Li, H.W. Tseng, D.C. Ralph and R.A. Buhrman, Science **336**, 555 (2012).

27. C-F. Pai, L. Liu, Y. Li, H.W. Tseng, D.C. Ralph and R.A. Buhrman, Appl. Phys. Lett. **101**, 122404 (2012).

28. J-C. Rojas-Sanchez, L. Villa, G. Desfonds, S. Gambarelli, J.P. Attane, J.M. De Teresa, C. Magen and A. Fert, Nat. Commun. **4**, 3944 (2013).

29. M. Matsushima, Y. Ando, S. Dushenko, R. Ohshima, R. Kumamoto, T. Shinjo and M. Shiraishi, Appl. Phys. Lett. **110**, 072404 (2017).

30. M.B. Jungfleisch, W. Zhang, J. Sklenar, W. Jiang, J.E. Pearson, J.B. Kettelson and A. Hoffmann, Phys. Rev. B **93**, 224419 (2016).

31. C. Zucchetti, M.-T. Dau, F. Bottegoni, C. Vergnaud, T. Guillet, A. marty, C. Beigne, S. Gambarelli, A. Picone, A. Calloni, G. Bussetti, A. Brambilla, L. Duo, F. Ciccacci, P.K. Das, J. Fujii, I. Vobornik, M. Finazzi and M. Jamet, Phys. Rev. B **98**, 184418 (2018).

32. A. Tulapurkar, Y. Suzuki, A. Fukushima, H. Kubota, H. Maehara, K. Tsunekawa, D.D. Djayaprawira, and S. Yuasa, Nature **438**, 339 (2005).

33. J.C. Rojas Sanchez J.-C. Rojas-Sánchez, N. Reyren, P. Laczkowski, W. Savero, J.-P. Attané, C. Deranlot, M. Jamet, J.-M. George, L. Vila, and H. Jaffrès, Phys. Rev. Lett. **112**, 106602 (2014).

34. S. Hirayama, S. Mitani, Y. Ohtani and S. Kasai, Appl. Phys. Express **11**, 013002 (2018).

35. J.M. Shaw, H.T. Nembach, T.J. Silva and C.T. Boone, J. Appl. Phys. **114**, 243906 (2013).

36. F. Rortais, S. Lee, R. Ohshima, S. Dushenko, Y. Ando and M. Shiraishi, Appl. Phys. Lett. **113**, 122408 (2018).

37. A. J. Berger, E.R. Edwards, H.T. Nembach, O. Karis, M. Weiler and T.J. Silva, Phys. Rev. B **98**, 024402 (2018).

**Figure captions**

**FIG. 1.** (a) Schematic of sample stacking. The Bi thickness is varied systematically from 1.2 to 10 nm. (b) Schematic of ST-FMR measurement circuit. An external magnetic field is applied in plane and the angle $\theta$ is set to 45° or 225°.

**FIG. 2.** (a) ST-FMR spectra from the Bi (5 nm)/Py (5 nm) sample, electric current frequencies ranging from 7 to 14 GHz. The black solid lines are fits. (b) Frequency dependence of resonance field. The red dashed line is a fit to the Kittel equation. (c) Full width at half maximum of the ST-FMR spectra as a function of Bi thickness ($f$=10 GHz). The blue and red solid circles show the results with the external magnetic field applied at 45° and 225°, respectively.

**FIG. 3.** (a) Symmetric ($V_S = CS$) and antisymmetric ($V_A = CA$) components as a function of Bi thickness. (b) Spin-conversion efficiency $\eta$ as a function of Bi thickness.

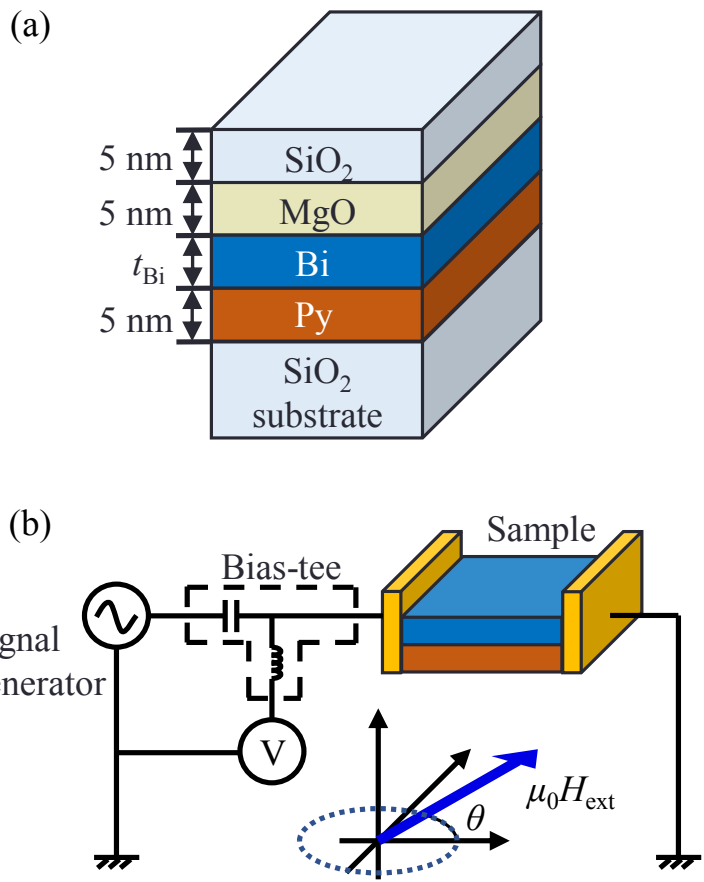

Fig. 1 Matsushima *et al*.

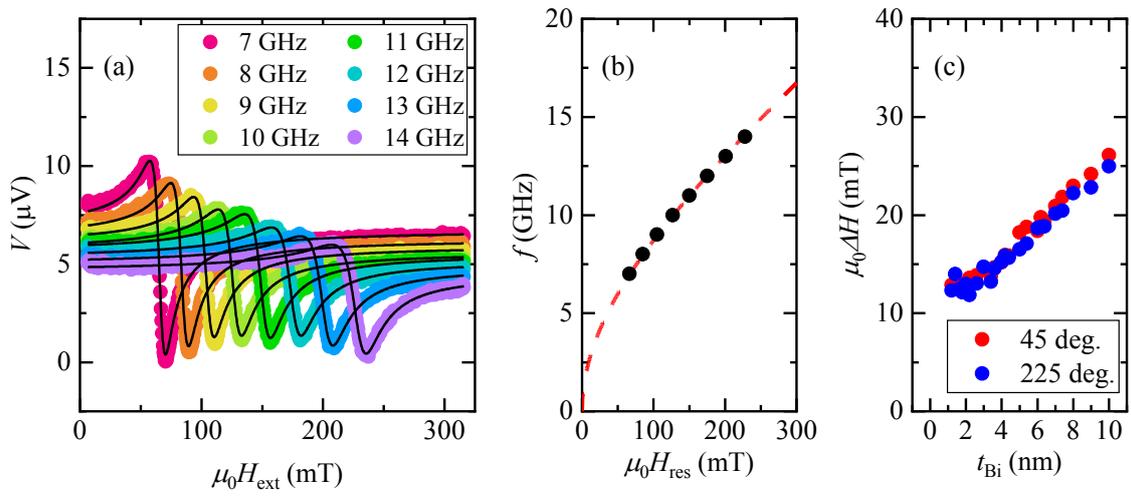

Fig. 2 Matsushima *et al*.

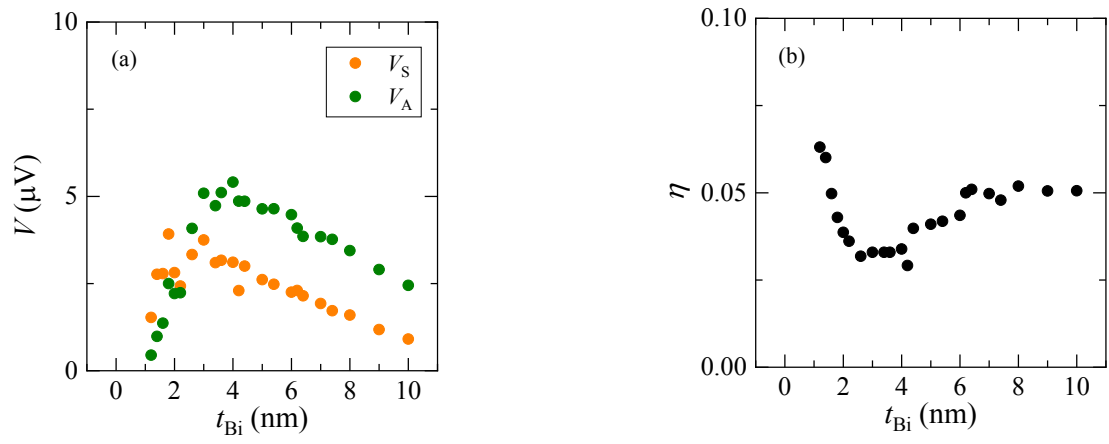

Fig. 3 Matsushima *et al*.